\documentclass[reprint,showpacs,amssymb,amsmath,aps,pra]{revtex4-1}
\usepackage{graphics}
\usepackage{bm}
\usepackage{amsmath}
\usepackage{amssymb}
\usepackage{amsthm}
\usepackage{graphicx}
\begin{document}

\title{Generalized conditional entropy optimization for qudit-qubit states}
\author{N. Gigena, R. Rossignoli}
\affiliation{Departamento de F\'{\i}sica-IFLP,
Universidad Nacional de La Plata, C.C. 67, La Plata (1900), Argentina}


\begin{abstract}
We derive a general approximate solution to the problem of minimizing the
conditional entropy of a qudit-qubit system resulting from a local measurement
on the qubit, which is valid for general entropic forms and becomes exact in
the limit of weak correlations. This entropy measures the average conditional
mixedness of the post-measurement state of the qudit, and its minimum among all
local measurements represents a generalized  entanglement of formation. In  the
case of the von Neumann entropy, it is directly  related to the quantum
discord. It is shown that at the lowest non-trivial order, the problem reduces
to the minimization of a quadratic form determined by the correlation tensor of
the system, the Bloch vector of the qubit  and the local concavity of the
entropy, requiring just  the diagonalization of a $3\times 3$ matrix. A simple
geometrical picture  in terms of an associated correlation ellipsoid is also
derived, which illustrates the link between entropy optimization and
correlation access and which is exact for a quadratic entropy. The approach
enables a simple estimation of the quantum discord. Illustrative results for
two-qubit states are  discussed.
\end{abstract}
\maketitle

\section{Introduction}
Quantification of quantum correlations in composite quantum systems is a topic
of great current interest \cite{Mo.12}. For pure states such correlations can
be identified with entanglement, which can be measured by the entropy of
entanglement \cite{BB.94}. Entanglement has been shown to be useful as a
resource for quantum teleportation \cite{CB.93} and pure state based quantum
computation \cite{JL.03,RBB.03}. For mixed states, however, the situation
becomes more complex and different measures have been introduced, such as the
entanglement of formation and the entanglement of distillation \cite{BD.96}.
Moreover, it has recently become clear that entanglement is not the only type
of non-classical correlation that a mixed quantum state can exhibit
\cite{Mo.12}. Most separable mixed states states, defined as convex mixtures of
product states \cite{WR.89}, can still possess a non-zero value of the quantum
discord \cite{OZ.01,HV.01,Zu.03}, defined as the minimum difference between two
quantum versions of the classical mutual information, or equivalently, the
classical conditional entropy \cite{OZ.01}. And a finite discord has been shown
to be present \cite{DSC.08} in the mixed state based algorithm of Knill and
Laflamme \cite{KL.98}, able to achieve an exponential speed up over the
classical algorithm with vanishing entanglement \cite{DFC.05}. Since then, several
other  measures of non-classical correlations for mixed states, sharing common
basic properties with the quantum discord, were introduced
\cite{Mo.12,Mo.10,DVB.10,RCC.10,SKB.11,BB.12,GA.12,POS.13,TM.13,GTA.13,NPA.13},
and various operational implications of discordant states have  been
provided \cite{Mo.12,SKB.11,GTA.13,NPA.13,PGA.11,AA.14}.

Entropy optimization is a central feature in many of these measures. In
particular, the quantum discord for a bipartite system requires the
minimization of the von Neumann conditional entropy  obtained as a result of a
local measurement on one of its components, over all such measurements, which
turns its evaluation difficult  (recently shown to be NP-complete \cite{YH.14})
in the general case. This conditional entropy is also interesting by itself,
since it measures the average conditional mixedness of the unmeasured component
after a measurement on the other. For pure states, this conditional entropy
vanishes for {\it any} local measurement based on rank one projectors, as the
post-measurement state will be pure and separable.  The optimization problem
arises then only for mixed states, for which the degree of mixedness of the
unmeasured side depends on the measurement performed on the other side. In
addition, its minimum represents the entanglement of formation between the
unmeasured component and a third partner  purifying the whole system
\cite{KW.04}.

In a previous work \cite{GR.14} we have analyzed the general properties of this
measurement dependent conditional entropy for {\it general} entropic forms.
This allows, in particular, to consider simple entropies like the so-called
linear entropy (a quadratic form in  the state $\rho$), which is directly
related to the purity and  whose minimization in a qudit-qubit system for
measurements on the qubit can be exactly determined \cite{GR.14}. In this work
we first provide a clear geometric picture of the optimization problem in a
qudit-qubit system in terms of the {\it correlation ellipsoid}, which
represents the set of post-measurement states of the unmeasured side and
depends on the correlation tensor $C$ of the system and the reduced state of
the qubit. It is shown that the exact optimization of the quadratic entropy
directly follows the largest semi-axis of this ellipsoid, maximizing
correlation access.

We then extend this approach to a general entropic form, deriving a quadratic
(in $C$) approximation to the conditional entropy valid for  a sufficiently
small correlation ellipsoid. The optimization problem becomes then equivalent
to the minimization of a $3\times 3$ quadratic form, being thus exactly
solvable and similar to that for the quadratic entropy with an {\it effective}
correlation tensor which takes into account the local concavity of the entropy.
The formalism is then applied to derive a  quadratic (in $C$) approximation to
the quantum discord, exact in the limit of weak correlations. Illustrative
results for two-qubit $X$ states are provided, which show the validity of the
present approach even beyond the very weak correlation limit.

\section{Formalism}

\subsection{Generalized conditional entropy after a local measurement}

We consider a bipartite  quantum state $\rho_{AB}$ with marginal states
$\rho_{A(B)}={\rm Tr}_{B(A)}\rho_{AB}$. We assume a measurement is performed on
system $B$, defined by a set of operators $M_j=I_A\otimes M^B_j$, such that the
operators $\Pi_j=M^{\dagger}_jM_j=I_A\otimes\Pi^B_j$ satisfy $\sum_j
\Pi_j=I_A\otimes I_B$. We then introduce the generalized conditional entropy
\cite{GR.14}
\begin{equation}
S_f(A|B_{\{\Pi_j\}})=\sum_{j}p_j S_f(\rho_{A/\Pi_j}) \label{EC}\,,
\end{equation}
where $p_j={\rm Tr}\,\rho_{AB}\,\Pi_j$ is the probability of outcome $j$,
$\rho_{A/\Pi_j}=({\rm Tr}_B\,\rho_{AB}\Pi_j)/p_j$ is the reduced state of $A$
after such outcome and
\begin{equation}
 S_f(\rho)={\rm Tr}\, f(\rho)\,,\label{Sf}\end{equation}
is a generalized entropic form \cite{CR.02}. Here
$f:[0,1]\rightarrow\mathbb{R}$ is a smooth strictly concave function satisfying
$f(0)=f(1)=0$, such that $S_f(\rho)\geq 0$, vanishing just for pure states.
Moreover, Eq.\ (\ref{Sf}) is then also strictly concave:  $S_f(\sum_\alpha
q_\alpha \rho_\alpha)\geq \sum_\alpha q_\alpha S_f(\rho_\alpha)$ if
$q_\alpha>0$, $\sum_\alpha q_\alpha=1$, with equality iff all $\rho_{\alpha}$
are equal \cite{pr,Bh.97}. This implies $S_f(\rho)\geq S_f(\rho')$ if
$\rho\prec\rho'$, i.e., if $\rho$ is {\it more mixed} than $\rho'$
\cite{CR.02,Bh.97}, entailing that $S_f(\rho)$ is maximum for $\rho$ maximally
mixed ($\rho=I/{\rm Tr}\,I$). We will set the normalization $2f(1/2)=1$, such
that $S_f(\rho)=1$ for a maximally mixed single qubit state, and assume
$f''(p)<0$  $\forall$ $p\in(0,1)$.

Eq.\ (\ref{EC}) is then a measure of the average conditional mixedness of the
state of $A$ after a measurement at $B$, and is non-negative. For
$f(p)=-p\log_2 p$, $S_f(\rho)$ is  the von Neumann entropy $S(\rho)$ and Eq.\
(\ref{EC}) becomes the conditional entropy introduced in the definition of
quantum discord \cite{OZ.01} (sec.\ \ref{D}). Generalizations of the
measurement independent von Neumann conditional entropy
$S(\rho_{AB})-S(\rho_B)$ (which is negative for pure entangled states) have
also  been recently considered \cite{ML.13,RA.13,AR.14}.

The concavity of $S_f(\rho)$ leads to general properties of Eq.\ (\ref{EC})
\cite{GR.14}. First, Eq.\ (\ref{EC}) cannot be greater than the entropy of the
marginal state of $A$: Since $\sum_j p_j\rho_{A/\Pi_j}= \rho_A$,
$S_f(\rho_A)=S_f(\sum_j p_j\rho_{A/\Pi_j})\geq \sum_j p_j S_f(\rho_{A/\Pi_j})$,
i.e.,
\begin{equation}
S_f(A)\geq S_f(A|B_{\{\Pi_j\}}), \label{concavity}
\end{equation}
with equality iff all $\rho_{A/\Pi_j}$ with $p_j>0$ are equal \cite{pr} (as
occurs for $\rho_{AB}=\rho_A\otimes\rho_B$). A measurement at $B$ cannot then
increase, on average, the mixedness of the state of $A$, for any choice of
measure $S_f$ used to quantify it.

Secondly, Eq.\ (\ref{EC}) is also concave: if $\rho_{AB}=\sum_\alpha
q_\alpha\rho_{AB}^{\alpha}$, with $q_\alpha>0$, $\sum_\alpha q_\alpha=1$, then
\cite{GR.14}
\begin{equation}
S_f(A|B_{\{\Pi_j\}})\geq\sum_{\alpha}q_{\alpha}S_f(A^{\alpha}|B^{\alpha}_{\{\Pi_j\}}),
\end{equation}
where $S_f(A^{\alpha}|B^{\alpha}_{\{\Pi_j\}})
=\sum_{j}p_j^{\alpha}S_f(\rho^{\alpha}_{A/\Pi_j})$ and $p_j^{\alpha}={\rm
Tr}\,\rho_{AB}^\alpha \Pi_j$. Uncertainty about  $A$ cannot then decrease with
state mixing. Furthermore, Eq.\ (\ref{EC}) cannot increase if a more detailed
measurement is performed: If $\Pi_j=\sum_k r_j^k\tilde\Pi_k$, where $r_j^k \geq
0$ and $\tilde\Pi_k=I_A\otimes\tilde\Pi_k^B$ are positive operators
representing a more detailed measurement ($\sum_k\tilde\Pi_k^B=I_B$, $\sum_j
r_j^k=1$), $\rho_{A/\Pi_j}=\sum_k p_j^{-1}r_j^k q_k \rho_{A/\tilde\Pi_{k}}$,
with $q_k={\rm Tr} \rho_{AB}\tilde\Pi_k$, $p_j=\sum_k r_j^k q_k$, and
\begin{equation}
S_f(A|B_{\{\Pi_j\}})\geq\sum_k q_k S_f(\rho_{A/\tilde\Pi_k})=S_f(A|B_{\{\tilde\Pi_k\}}).
\end{equation}
Conditional entropy minimization is therefore achieved with measurements based
on rank one projectors $\tilde{\Pi}_k^B$. In the case of pure states
$\rho_{AB}^2=\rho_{AB}$, the conditional entropy (\ref{EC}) vanishes in fact
for any measurement based on rank-one projectors, as $\rho_{A/\tilde{\Pi}_k}$
will be pure \cite{GR.14}.

If $C$ is a system purifying  $A+B$, such that $\rho_{AB}={\rm
Tr}_C|\Psi_{ABC}\rangle\langle \Psi_{ABC}|$, the minimum conditional entropy
among all local measurements at $B$ is the  generalized entanglement of
formation between $A$ and $C$ \cite{KW.04,GR.14,MD.11}:
\begin{equation}
\mathop{\rm Min}_{\{\Pi_j\}}S_f(A|B_{\{\Pi_j\}})=E_f(A,C)\,,\label{ECm}
\end{equation}
where $E_f(A,C)$ is the convex roof extension of the generalized entanglement
entropy of pure states ($E_f(A,C)=S_f(\rho_A)=S_f(\rho_C)$ if
$\rho_{AC}=\rho_{AC}^2$). It is an entanglement monotone \cite{Vi.00}.

\subsection{The qudit-qubit case and its geometrical picture}
\subsubsection{General expressions}
Let us now assume that  $B$ is a single qubit, with $A$ a  system with Hilbert
space dimension $d_A$ (qudit). We can describe a general state of this system
in terms of the Pauli operators
$\bm{\sigma}_B=(\sigma_{x},\sigma_{y},\sigma_{z})$ for system $B$ and an
analogous set  of $D_A=d_A^2-1$ orthogonal hermitian operators $\bm{\sigma}_A$
for system $A$, satisfying (for $\mu,\mu'=1,\ldots,D_A$)
\begin{equation}{\rm Tr}\,\sigma_{A\mu}=0\,,\;\;{\rm
Tr}\,\sigma_{A\mu}\sigma_{A\mu'}=d_A\delta_{\mu\mu'}\,.\label{sga}
\end{equation}

In the generalized Fano-Bloch representation \cite{F.55}, an arbitrary state of
this system can be written as
\begin{equation}
\rho_{AB}=\rho_A\otimes\rho_B+{\frac{1}{2d_A}}\sum_{\mu,\nu}C_{\mu\nu}\,
\sigma_{A\mu}\otimes\sigma_{B\nu}\,, \label{rhoab}
\end{equation}
where $\rho_{A(B)}$ are the reduced states
 \begin{equation}
\rho_{A}=\frac{1}{d_A}(I_A+\bm{r}_A\cdot\bm{\sigma}_A),
 \;\;\rho_B=\frac{1}{2}(I_B+\bm{r}_{B}\cdot\bm{\sigma}_{B}) \,,\end{equation}
with $\bm{r}_{A(B)}=\langle \bm{\sigma}_{A(B)}\rangle\equiv{\rm
Tr}\,\rho_{A(B)}\bm{\sigma}_{A(B)}$, and
\begin{equation}C_{\mu\nu}=\langle\sigma_{A\mu}\otimes\sigma_{B\nu}\rangle
 - \langle\sigma_{A\mu}\rangle\langle\sigma_{B\nu}\rangle
\,,\label{Cab}
\end{equation}
are the elements of the \emph{correlation tensor} $C$ of the system, which form
a real $D_A\times 3$ matrix. $C$ may be seen as an object analogous to an
inertia tensor, in the sense that for a unit vector $\bm{k}$ in $\mathbb{R}^3$,
the number $|C\bm{k}|$ is a measure of the amount of correlations for spin
direction $\bm{k}$ at $B$. Through its singular value decomposition
 \begin{equation}
 C=UDV^T\,,\;\;D_{\mu\nu}=\delta_{\mu\nu}C_\mu\,,\label{SVD}
 \end{equation}
where $U$, $V$ are real orthonormal $D_A\times D_A$ and $3\times 3$ matrices
($U^T=U^{-1}$, $V^T=V^{-1}$, $T$ indicating transpose) and $C_\mu^2$ the
eigenvalues of the $3\times 3$ matrix $C^TC$ (identical with the non-zero
eigenvalues of $CC^T$), we may always select orthogonal operators
$\tilde{\sigma}_{A\mu}=\sum_{\mu'}U_{\mu'\mu}\sigma_{A\mu'}$,
$\tilde{\sigma}_{B\nu}=\sum_{\nu'}V_{\nu'\nu}\sigma_{B\nu'}$ satisfying Eqs.\
(\ref{sga}), such that just
 three operators in $A$ will be connected through $C$ with those of $B$:
\begin{equation}\sum_{\mu,\nu}C_{\mu\nu}\,\sigma_{A\mu}\otimes \sigma_{B\nu}=
\sum_{\mu=1}^3 C_\mu\tilde{\sigma}_{A\mu}\otimes\tilde{\sigma}_{B\mu}\,.
\label{CD}\end{equation}

A projective measurement on qubit $B$ is characterized by the measurement
operators $\Pi_{\pm\bm{k}}^B=\frac{1}{2}(I\pm\bm{k}\cdot\bm{\sigma}_{B})$,where
$\bm{k}$ is a unit vector in $\mathbb{R}^3$. After this measurement is
performed, the reduced state of $A$ and its probability are
\begin{eqnarray}
\rho_{A/\Pi_{\pm\bm{k}}}&=&\rho_A\pm\frac{1}{d_A}
\left(\frac{C\bm{k}}
{1\pm\bm{r}_B\cdot\bm{k}}\right)\cdot\bm{\sigma}_A\,,\label{rhak}\\
p_{\pm\bm{k}}&=&{\textstyle\frac{1}{2}}(1\pm\bm{r}_B\cdot\bm{k}),\label{PME}
\end{eqnarray}
implying that the Bloch vector characterizing the post-measurement state of $A$ is
\begin{equation}\bm{r}_{A/\pm\bm{k}}=\bm{r}_A\pm\frac{C\bm{k}}{1\pm\bm{r}_B\cdot\bm{k}}
\,.\label{rak}
\end{equation}
The ensuing conditional entropy $S_f(A|B_{\bm{k}})\equiv
S_f(A|B_{\{\Pi_{\bm{k}},\Pi_{-\bm{k}}\}})$ becomes
\begin{equation}
S_f(A|B_{\bm{k}})=\sum_{\nu=\pm 1}p_{\nu\bm{k}}
S_f(\rho_{A/\Pi_{\nu\bm{k}}})\,,
\label{ECk}
\end{equation}
with $S_f(\rho_{A/\Pi_{\nu\bm{k}}})=\sum_{i=1}^{d_A}f(p^A_{i/\nu\bm{k}})$ and
$p^A_{i/\pm\bm{k}}$ the eigenvalues of $\rho_{A/\pm\Pi_{\bm{k}}}$.  For a
general POVM measurement $M_B$ based on a set of rank one operators
$\sqrt{r_{\bm{k}}}\,\Pi_{\bm{k}}^B$, with $\sum_{\bm
k}r_{\bm{k}}\Pi_{\bm{k}}^{B}=I_B$,  we should just replace (\ref{ECk}) by
\begin{equation}
S_f(A|B_{\{r_{\bm k}\Pi_{\bm k}\}})=\sum_{\bm{k}}r_{\bm{k}}p_{\bm{k}}
S_f(\rho_{A/\Pi_{\bm{k}}})\,.
\label{ECk2}
\end{equation}

\subsubsection{Geometrical Picture}
\begin{figure}[htp]
\centering
\includegraphics[scale=0.5]{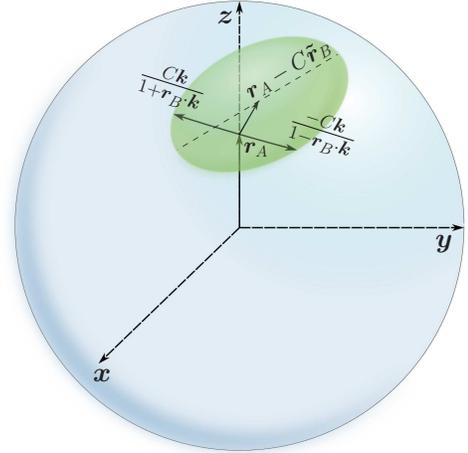}
\caption{(Color online) Schematic representation of the correlation ellipsoid
(\ref{elA}) depicting the possible Bloch vectors $\bm{r}_{A/\bm{k}}$ of the
post-measurement state of $A$. It is centered at $\bm r_A -C\tilde{\bm r}_B$,
with $\tilde{\bm r}_B=\bm{r}_B/(1-r_B^2)$. For a given direction $\bm k$ on the
unit sphere  of $B$, the vectors $\bm r_{A/\pm\bm k}$ in $A$ are the endpoints
of a chord running through $\bm r_A$. If $\bm{r}_B=\bm{0}$, the ellipsoid
becomes centered at $\bm{r}_A$. } \label{f1}
\end{figure}

The set of all post-measurement vectors (\ref{rak}) will form in general a three
dimensional ellipsoid, which we will denote as {\it correlation ellipsoid}
(Fig.\ \ref{f1}). If $\bm{r}_B=\bm{0}$  ($\rho_B$ maximally mixed),
$\delta\bm{r}_{A}=\bm{r}_{A/\bm{k}}-\bm{r}_A=C\bm{k}$,  and the ellipsoid will
be centered at $\bm{r}_A$. Its principal axes will lie along the principal
directions associated with the operators $\tilde{\sigma}_{A\mu}$ in (\ref{CD}),
and their lengths will be the singular values $C_\mu$.

For general values of $\bm{r}_B$, defining first
$\tilde{\bm{k}}=\frac{\bm{k}}{1+\bm{r}_B\cdot\bm{k}}$, such that
$1-\bm{r}_B\cdot\tilde{\bm k}=\frac{1}{1+\bm{r}_B\cdot\bm{k}}$, the
unit sphere $\bm{k}\cdot\bm{k}=1$ is seen to map into the shifted ellipsoid
$\tilde{\bm{k}}\cdot\tilde{\bm{k}}=(1-\bm{r}_B\cdot\tilde{\bm{k}})^2$, which
can be written explicitly as
\begin{eqnarray}
&&\left(\tilde{\bm k}+\tilde{\bm r}_B\right)^T (1-r_B^2)N_B
\left(\tilde{\bm k}+\tilde{\bm r}_B\right)=1\,,\label{kte}\\
 &&N_B=I-\bm{r}_B\bm{r}_B^T\,,\label{NB}\end{eqnarray}
where $r_B=|\bm{r}_B|$, $\tilde{\bm r}_B=\bm{r}_B/(1-r_B^2)$ and $N_B$
is a $3\times 3$ matrix (positive definite if $r_B<1$). This ellipsoid has
eccentricity $\bm{r}_B$, with the origin as one of its foci. Next, $C$ in
(\ref{rak}) will map Eq.\ (\ref{NB}) into a shifted ellipsoid centered at
$\bm{r}_A-C\tilde{\bm r}_B$:
\begin{equation}
\left(\delta\bm{r}_{A}+C\tilde{\bm r}_B\right)^T(1-r_B^2)(CN_B^{-1}C^T)^{-1}
 \left(\delta\bm{r}_{A}+C\tilde{\bm r}_B\right)=1\,, \label{elA}
 \end{equation}
where  $CN_B^{-1}C^T$ is a positive semidefinite matrix (its inverse in
(\ref{elA}) is taken within the subspace associated with the operators
$\tilde{\sigma}_{A\mu}$ in (\ref{CD})). The principal axes of this ellipsoid
are determined by its eigenvectors $\bm{k}^A_\mu$, i.e.,
\begin{equation} CN_B^{-1}C^T\bm{k}_\mu^A=\lambda_\mu\bm{k}_\mu^A\,,
\label{CNCeig}\end{equation}
associated with the non-zero eigenvalues $\lambda_\mu$,  with the semi-axes
lengths given by $\sqrt{\lambda_\mu/(1-r_B^2)}$.

For pure states $\rho_{AB}^2=\rho_{AB}$, $\rho_{A/\Pi_{\bm{k}}}$ is pure
$\forall$ $\bm{k}$, so that $|\bm{r}_{A/\bm{k}}|^2=d_A-1$ $\forall$ $\bm{k}$.
For instance, in a two-qubit system, by suitable choosing the local $x,y,z$
axes, the Schmidt decomposition allows to write any pure state as
$|\Psi_{AB}\rangle=\sqrt{p}|00\rangle+\sqrt{1-p}|11\rangle$. This leads to
$C_{\mu\nu}=\delta_{\mu\nu} C_\mu$ and $r_{A\mu}=r_{B\mu}=\delta_{\mu z}r_B$,
with $C_x=-C_y=2\sqrt{p(1-p)}$, $r_B=2p-1$ and $C_z=1-r_B^2=C_x^2$. It is then
verified that for $p\in(0,1)$, the ellipsoid (\ref{elA}) becomes the Bloch
sphere of $A$ ($C N_B^{-1}C^T=(1-r_B^2)I$, $C\tilde{\bm r}_B=\bm{r}_A$).

\subsection{The case of the quadratic entropy}
\subsubsection{Explicit expressions and minimum conditional entropy}
The evaluation of $S_f(\rho_A)$ for a general $f$ requires the eigenvalues of
$\bm{r}_A\cdot\bm{\sigma}_A$. However, in the case of the quadratic entropy
\begin{equation} S_2(\rho)=2(1-{\rm Tr}\,\rho^2)\,,\label{S2}\end{equation}
obtained for $f(p)=p(1-p)$ (also denoted as linear
entropy as it follows from  the approximation $-\ln p \approx 1-p$
in the von Neumann
entropy), a close evaluation in terms of $|\bm{r}_A|$ becomes feasible. We
obtain, using Eq.\ (\ref{sga}),
\begin{equation}S_2(\rho_A)=2\left(1-\frac{1+|\bm{r}_A|^2}{d_A}\right)\,.
\label{S2A}\end{equation}
Eq.\ (\ref{S2A}) is trivially related to the {\it purity} ${\rm
Tr}\,\rho_A^2=(1+|\bm{r}_A|^2)/d_A$  and to the standard squared distance to
the maximally mixed state, $||\rho_A-I_A/d_A||^2=|\bm{r}_A|^2/d_A$, where
$||O||^2={\rm Tr}\,O^\dagger O$. Eq.\ (\ref{S2A})  shows that $|\bm{r}_A|^2\leq
d_A-1$, with $|\bm{r}_A|^2=d_A-1$ just for pure states $\rho_A^2=\rho_A$.

Using Eqs.\ (\ref{PME}), (\ref{rak})  and (\ref{S2A}), the conditional entropy
(\ref{ECk}) in the quadratic case can be
expressed as \cite{GR.14}
\begin{eqnarray}
S_2(A|B_{\bm{k}})&=&S_2(\rho_A)-\Delta S_2(A|B_{\bm k})\,,\label{S2k}\\
\Delta S_2(A|B_{\bm{k}})&=&\frac{2}{d_A}\frac{|{C}\bm{k}|^2}{1-(\bm{r}_B
\cdot\bm{k})^2}=\frac{2}{d_A}\frac{\bm{k}^TC^TC\bm{k}}
{\bm{k}^T N_B\bm{k}}\,,\label{S2x}
\end{eqnarray}
where $C^TC$ and $N_B$ (Eq.\ (\ref{NB}))  are $3\times 3$ positive
semi-definite matrices. The entropy decrease  (\ref{S2x}) is then non-negative
and represents the average conditional purity gain due to the measurement on
$B$. It is independent of $\bm{r}_A$.

Since Eq.\ (\ref{S2x}) is a ratio of quadratic forms, the  direction $\bm{k}$
leading to the maximum entropy decrease can be obtained by solving the weighted
eigenvalue problem \cite{GR.14}
\begin{equation}
C^TC\bm{k}=\lambda N_B\bm{k}\,,\label{eig2}
\end{equation}
which implies ${\rm Det}[C^TC-\lambda N_B]=0$, and selecting the eigenvector
$\bm{k}$ associated with the largest eigenvalue $\lambda_{\rm max}$. This leads
to $\Delta S_2(A|B_{\bm k})\leq 2\lambda_{\rm max}/d_A$ $\forall$ $\bm{k}$,
i.e.,
\begin{equation}
\mathop{\rm Min}_{\bm{k}}S_2(A|B_{\bm{k}})=S_2(\rho_A)-\frac{2}{d_A}\lambda_{\rm max}\,.
\label{S2min}
\end{equation}
We may also express (\ref{S2x}) as the  quadratic form
\begin{equation}
\Delta S_2(A|B_{\bm{k}})=\frac{2}{d_A}\bm{k}_N^TC_N^TC_N\bm{k}_N
\,,\;\;C_N=CN_B^{-1/2}\,,\label{S2xx}
\end{equation}
where $\bm{k}_N=N_B^{1/2}\bm{k}/|N_B^{1/2}\bm{k}|$ is a unit vector.  Eq.\
(\ref{eig2}) is in fact equivalent to $C_N^TC_N\bm{k}_N=\lambda\bm{k}_N$,
showing that $\sqrt{\lambda_{\rm max}}$ is the {\it maximum singular value} of
$C_N$.

An important final remark is that for this entropy,  generalized (POVM)
measurements on qubit $B$ {\it cannot decrease the projective minimum (\ref{S2min})}.
\nopagebreak
\begin{proof}
For a measurement based on rank one operators
$\sqrt{r_{\bm{k}}}\,\Pi_{\bm{k}}^B$,
$\Pi_{\bm{k}}^B=\frac{1}{2}(I_B+\bm{k}\cdot\bm{\sigma}_B)$, with $\sum_{\bm
k}r_{\bm{k}}\Pi^{B}_{\bm{k}}=I_B$, Eq.\ (\ref{ECk2}) leads, for $\Delta
S_2(A|B_{M})\equiv S_2(\rho_A)-S_2(A|B_{M})$, to
\begin{eqnarray}\Delta S_2(A|B_{\{r_{\bm k}\Pi_{\bm k}\}})&=&{\frac{1}{d_A}}
\sum_{\bm{k}}r_{\bm k}\frac{|C\bm{k}|^2}{1+\bm{r}_B\cdot\bm{k}}\nonumber\\
&=&\frac{1}{d_A}\sum_{\bm{k}}r_{\bm k}(1-\bm{r}_B\cdot\bm{k})
\frac{|C\bm{k}|^2}{1-(\bm{r}_B\cdot\bm{k})^2}\nonumber\\
&\leq &\frac{\lambda_{\rm max}}{d_A}\sum_{\bm{k}}r_{\bm{k}}(1-\bm{r}_B\cdot\bm{k})=
\frac{2}{d_A}\lambda_{\rm max}\label{pvm}\,,
\end{eqnarray}
where we used Eq.\ (\ref{S2min}). This ensures that the lowest conditional
entropy (maximum $\Delta S_2(A|B_{M})$) is reached for the projective
measurement determined by Eq.\ (\ref{eig2}).
\end{proof}

\subsubsection{Geometrical picture of optimum measurement}

\begin{figure}[htp]
\includegraphics[scale=.8]{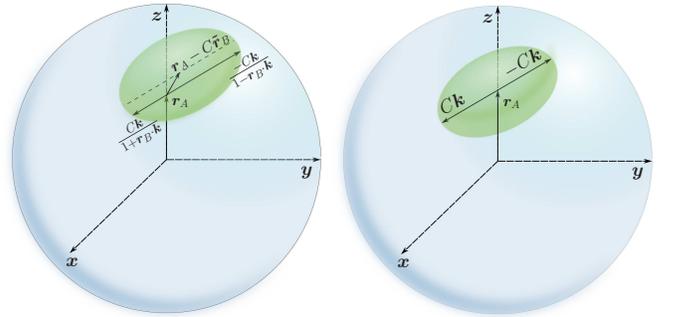}
\caption{(Color online) Bloch vectors of the post-measurement states of
$\bm{A}$ that minimize the quadratic conditional entropy (\ref{S2k}). The left
panel depicts the general case, whereas the right panel the case
$\bm{r}_B=\bm{0}$. The increase $\delta{\bm r}_{A}=\bm{r}_{A/\bm{k}}-\bm{r}_A$
is parallel to the largest semiaxis of the correlation ellipsoid, and coincides
with it when $\bm{r}_B=\bm{0}$.} \label{f2}
\end{figure}

Eq.\ (\ref{eig2}) is also the counterpart at $B$ of the eigenvalue Eq.\
(\ref{CNCeig}) (equivalent to $C_NC_N^T\bm{k}^A=\lambda\bm{k}^A$), which
determined the correlation ellipsoid axes, having both {\it the same} non-zero
eigenvalues $\lambda_\mu$, with related eigenvectors ($C^TC\bm{k}=\lambda
N_B\bm{k}$ $\Rightarrow$ $CN_B^{-1}C^T\bm{k}^A=\lambda \bm{k}^A$ for
$\bm{k}^A\propto C\bm{k}$). Hence, the optimizing measurement of the quadratic
entropy is precisely that leading to $\delta \bm{r}_{A}\propto C\bm{k}$ {\it
parallel to the major semi-axis of the correlation ellipsoid} (Fig.\ \ref{f2}).

If  $\bm{r}_B=\bm{0}$, the Bloch vector of post-measurement state of $A$ is
just $\bm{r}_{A/\pm\bm{k}}=\bm{r}_A\pm C\bm{k}$, with equal probabilities for
$\bm{k}$ and $-\bm{k}$, and the correlation ellipsoid becomes  centered at
$\bm{r}_A$ (right panel in Fig.\ \ref{f2}). Hence, for a given direction
$\bm{k}$, the two possible post-measurement Bloch vectors are located
diametrically opposite on this ellipsoid. The vector $\bm{k}$ optimizing the
quadratic entropy leads then to $\delta\bm{r}_{A}=\pm C\bm{k}$ directly {\it
coincident} with the major semi-axis, with $\lambda_{\rm max}={\rm max}_\mu
\{C_\mu^2\}$, representing its squared length. Note that in this case $N_B=I$
and Eq.\ (\ref{eig2}) becomes just $C^TC\bm{k}=\lambda\bm{k}$. Hence, the
optimizing $\bm{k}$ leads to {\it maximum  correlation}:
$|C\bm{k}|=\sqrt{\bm{k}^T C^TC\bm{k}}=\sqrt{\lambda_{\rm max}}$, with
$|C\bm{k}'|\leq |C\bm{k}|$ for any other direction $\bm{k}'$.

Since the conditional entropy is a measure of the average uncertainty about $A$
as a result of a measurement on $B$, its minimization implies making use of the
maximum amount of correlations available by a measurement on $B$. If the
correlation tensor measures the spatial distribution of correlations, the
measurement that maximizes correlations access should be in principle that
leading to a maximum length of $C\bm{k}$, which is precisely the measurement
minimizing the quadratic conditional entropy.

For $\bm{r}_B\neq \bm{0}$, the effect of $N_B^{-1}$ in Eq.\ (\ref{elA}) is to
deform the $\bm{r}_B=\bm{0}$ correlation ellipsoid, expanding it  along the
direction of $C\bm{r}_B$. Accordingly, in Eqs.\ (\ref{S2x})--(\ref{S2xx}) $N_B$
will favor measurements with $\bm{k}$ along or close  to $\bm{r}_B$, i.e., in
the basis  of ${\rho_B}$'s eigenstates. In order to understand this result,
note that for $\bm{r}_B\neq\bm{0}$,  $C$ in Eq.\ (\ref{rak}) acts on vectors
$\tilde {\bm k}_{\pm}={\pm\bm k}/({1\pm\bm r_B\cdot\bm k})$ which have a
direction dependent norm and lie on the surface of the shifted ellipsoid
(\ref{kte}), making correlation access dependent not only on $C$ but also on
$\bm{r}_B$. Nonetheless, it is seen from Eq.\ (\ref{kte}) that vectors
$N_B^{1/2}\tilde{\bm k}_{\pm}$  lie on a shifted sphere, forming a chord that
passes through the origin. The origin will divide this chord in two segments
whose length's product is $|N_B^{1/2}\tilde{\bm k}_+||N_B^{1/2}\tilde{\bm
k}_-|=\frac{\bm{k}^TN_B\bm{k}}{1-(r_B\cdot\bm k)^2}=1$.

Since $C\tilde{\bm{k}}_{\pm}=C_N N_B^{1/2}\tilde{\bm{k}}_{\pm}$ (Eq.\
(\ref{S2xx})),  the ellipsoid (\ref{elA}) may be seen as the image of the
previous sphere under the linear transformation $C_N$. As before, if $C_N$
measures the effective spatial distribution of correlations,  the product
\[|C_N(N_B^{1/2}\tilde{\bm k}_+)||C_N(N_B^{1/2}\tilde{\bm k}_-)|=
 \frac{|C\bm{k}|^2}{1-(\bm{r}_B\cdot\bm{k})^2}\,,\]
which is just proportional to $\Delta S_2(A|B_{\bm{k}})$ (Eq.\ (\ref{S2x})), is
a measure  of correlations along direction $\bm k$ at $B$. The direction $\bm{k}$
that minimizes $S_2(A|B_{\bm{k}})$ is then  precisely that which
maximizes this product.

\subsection{Conditional entropy and optimal measurement in the weakly correlated limit}
We now discuss the main general result of this manuscript. We will extend the
previous results to a {\it general entropy} $S_f$, within the {\it weakly
correlated regime}. This regime refers to the case where the correlation
ellipsoid  (Fig.\ \ref{f1}) {\it is sufficiently small:}
$|\delta\bm{r}_{A}|=|\frac{C\bm{k}}{1\pm \bm{r}_B\cdot\bm{k}}|\ll 1$ $\forall$
$\bm{k}$ in  (\ref{rak}). In this situation, we may consider an expansion of
the conditional entropy (\ref{ECk}) around $\rho_A$, up to {\it second order}
in $\delta\rho_A=\delta\bm{r}_A\cdot\bm{\sigma}_A/d_A$.  The result is
\begin{equation}
S_f(A|B_{\bm{k}})\approx S_f(\rho_A)-\frac{2}{d_A}
\frac{\bm{k}^T C^T \Lambda_f(\rho_A)C\bm{k}}{\bm{k}^T N_B\bm{k}}
\label{Sfg}\,,
\end{equation}
where $N_B$ is the $3\times 3$ matrix (\ref{NB}) and $\Lambda_f(\rho_A)$
denotes a  scaled $D_A\times D_A$ Hessian matrix, of elements
\begin{eqnarray}
[\Lambda_f(\rho_A)]_{\mu\mu'}&=&\frac{1}{4d_A}\sum_{i,j}R_{ij}
\langle i|\sigma_{A\mu}|j\rangle\langle j|\sigma_{A\mu'}|i\rangle\,,\label{Hfg}\\
R_{ij}&=&{(1-\delta_{ij})\frac{f'(p^A_i)-f'(p^A_j)}{p^A_j-p^A_i}-\delta_{ij}f''(p_i^A)}
\,,\label{Hfg2}
\end{eqnarray}
where $\rho_A|i\rangle=p_i^A|i\rangle$. Actually, just the
$3\times 3$ submatrix of $\Lambda_f(\rho_A)$ corresponding to the three
principal directions selected by $C$ in Eq.\ (\ref{CD}), is actually required
in (\ref{Sfg}).

\begin{proof}
We start from the second order expansion of the eigenvalues $p_{i/\bm{k}}^A$
of the post-measurement state (\ref{rhak}),
\begin{equation}p^A_{i/\bm{k}}\approx p_i^A+\langle i|\delta \rho_A|i\rangle
+\sum_{j\neq i} \frac{|\langle j|\delta\rho_A|i\rangle|^2}{p^A_i-p^A_j}\,,
 \label{exp1}\end{equation}
where $p_i^A$ are those of $\rho_A$ and $\delta\rho_A=\frac{1}{d_A}\frac{\pm
C\bm{k}}{1\pm\bm{r}_B\cdot\bm{k}}\cdot\bm{\sigma}_A$. The ensuing second order
expansion of the entropy in (\ref{ECk}),
\begin{equation}S_f(\rho_{A/\Pi_{\bm{k}}})
\approx S_f(\rho_A)+\sum_i [f'(p^A_i)\delta p^A_{i}+\frac{1}{2}f''(p^A_i)
 \delta p^{A\;2}_{i}]\,,\label{exp2}\end{equation}
where $\delta p^A_i=p^A_{i/\bm{k}}-p_i^A$, leads then to Eqs.\ (\ref{Sfg})--(\ref{Hfg2}),
after using (\ref{exp1}) and neglecting higher order terms. Note that
$R_{ij}=-f''(p_{ij}^*)$, with $p_{ij}^*$ between $p_i^A$ and $p_j^A$, entailing
$R_{ij}>0$ $\forall$ $i,j$, with  $R_{ij}\rightarrow -f''(p_i)$ if
$p_j\rightarrow p_i$. If $p_i^A>0$ $\forall$ $i$, $R_{ij}$ is finite $\forall$
$i,j$ for any $f$ of the form considered.
\end{proof}
The positivity of $R_{ij}$  $\forall$ $i,j$ implies that $\Lambda_f(\rho_A)$ is
positive definite, and hence that $C^T\Delta_f(\rho_A)C$ is positive
semidefinite. The entropy decrease
\begin{equation}
\Delta S_f(A|B_{\bm k})=S_f(A)-S_f(A|B_{\bm k})\approx
\frac{2}{d_A}
\frac{\bm{k}^T C^T \Lambda_f(\rho_A)C\bm{k}}{\bm{k}^T N_B\bm{k}}
\label{deltaf}
\end{equation}
then remains non-negative in the present approximation.

In the case of the quadratic entropy, $R_{ij}=4$ $\forall$ $i,j$, and Eqs.\
(\ref{sga}) and (\ref{Hfg}) lead to  $\Lambda_2(\rho_A)=I$, reducing Eq.\
(\ref{Sfg}) to Eqs.\ (\ref{S2k})-(\ref{S2x}). On the other hand, for
$\bm{r}_A\rightarrow \bm{0}$ (maximally mixed $\rho_A$),  $p_i^A=1/d_A$
$\forall$ $i$ and $R_{ij}\rightarrow -f''(1/d_A)$ $\forall$ $i,j$, implying
that Eq.\ (\ref{Hfg}) becomes again proportional to  the identity matrix
$\forall$ $S_f$:
\begin{equation}
\Lambda_f(I_A/d_A)={\textstyle\frac{1}{4}}|f''(1/d_A)|\,I\label{Hra0}\,.
\end{equation}
Hence,  Eqs.\ (\ref{deltaf})--(\ref{Hra0}) lead to $\Delta S_f(A|B_{\bm{k}})
\propto\Delta_2(A|B_{\bm{k}})$ $\forall$ $S_f$. In this limit the measurement
minimizing $S_f(A|B_{\bm{k}})$ is then {\it universal}, i.e., the same as that
optimizing the quadratic entropy $\forall$ $S_f$.

In the general case, the matrix  (\ref{Hfg}) will introduce an additional
``anisotropy'',  which will depend on $\rho_A$ and the choice of $f$, and which
represents  the effect of the ``concavity excess'' of $S_f$ at $\rho_A$ in
comparison with that of the quadratic entropy.  Nonetheless, Eq.\ (\ref{Sfg})
shows that in the weakly correlated regime, $S_f(A|B_{\bm{k}})$ becomes
equivalent to the quadratic conditional entropy (\ref{S2k}) for an effective
``deformed'' correlation tensor
\begin{equation}C_f=\sqrt{\Lambda_f(\rho_A)}\,C\,.\label{rep}\end{equation}

Minimization of Eq.\ (\ref{Sfg}) over $\bm{k}$ then leads again to a { $3\times
3$ weighted eigenvalue problem},
\begin{equation}
C^T \Lambda_f(\rho_A)C\bm{k}=\lambda_f N_B\bm{k}\,,\label{eigg}
\end{equation}
implying ${\rm Det}[C^T \Lambda_f(\rho_A)C-\lambda N_B]=0$.
The minimum is  obtained  for $\bm{k}$ along the direction of the eigenvector
associated with the largest eigenvalue $\lambda_{f {\rm max}}$ of (\ref{eigg}):
\begin{equation}
\mathop{\rm Min}_{\bm{k}}S_f(A|B_{\bm{k}})\approx S_f(\rho_A)-\frac{2}{d_A}
\lambda_{f {\rm max}}\,.\label{Sfmin}
\end{equation}
Moreover,  Eq.\ (\ref{deltaf}) can be rewritten as
\begin{equation}
\Delta S_f(A|B_{\bm{k}})\approx
\frac{2}{d_A}\bm{k}_N^T C_N^T\Lambda_f(\rho_A)C_N\bm{k}_N\,,
\label{Sfxx}
\end{equation}
with $C_N$ and $\bm{k}_N$ defined as in (\ref{S2xx}).

The geometric picture of these results is, therefore, similar to that for the
quadratic entropy, after replacing $C$ with  the deformed correlation tensor
(\ref{rep}). As in the quadratic case, in the approximation (\ref{exp2}) POVM
measurements will not decrease the projective minimum (\ref{Sfgmin}). The
argument is the same as that of Eq.\ (\ref{pvm}), after replacing $C$ with
$C_f$.

\subsection{The two-qubit case}
Let us now examine the case $d_A=2$. The  entropy  $S_f(\rho_A)$ of a general
single qubit state  $\rho_A=\frac{1}{2}(I_A+\bm{r}_A\cdot\bm{\sigma})$ will
depend just on the length of the Bloch vector $\bm{r}_A$:
\begin{equation}
S_f(\rho_A)=\sum_{\nu=\pm 1}f({\textstyle\frac{1+\nu |\bm{r}_A|}{2}})=h_f(|\bm{r}_A|)
 \label{Hdef},
\end{equation}
where $h_f(r)$ is a concave {\it strictly decreasing} function of $r$
for any strictly concave $f$.
The conditional entropy (\ref{ECk}) can  then be written as
\begin{equation}
S_f(A|B_{\bm{k}})=\sum_{\nu=\pm 1} p_{\nu\bm{k}}\,
h_f\left(\left|\bm{r}_A+\nu\frac{C\bm{k}}{1+\nu\bm{r}_B\cdot\bm{k}}\right|\right)\,,
\label{Sh(A|B)}
\end{equation}
 where $C$ is now a $3\times 3$ matrix.

If $\bm{r}_A=\bm{r}_B=\bm{0}$ (maximally mixed marginals), Eq.\ (\ref{Sh(A|B)})
reduces  to
\begin{equation}S_f(A|B_{\bm{k}})=h_f(|C\bm{k}|)\;\;\;\;(\bm{r}_A=\bm{r}_B=\bm{0})
\label{ra0}\,.\end{equation} Hence, in this case its minimum is reached, for
{\it any } $S_f$, for that $\bm{k}$ which maximizes  $|C\bm{k}|$, i.e.,
\begin{equation}
\mathop{\rm Min}_{\bm{k}}S_f(A|B_{\bm{k}})=h_f(\sqrt{\lambda_{\rm max}})
\;\;\;\;(\bm{r}_A=\bm{r}_B=\bm{0})\,,\label{ra02}
\end{equation}
where $\lambda_{\rm max}$ is the largest eigenvalue of $C^TC$ and $\bm{k}$ the
associated eigenvector ($\sqrt{\lambda_{\rm max}}=C_{\rm max}$ is the largest
singular value of $C$). Non-projective measurements will not decrease this
value, since $h_f(|C\bm{k}|)\geq h_f(\sqrt{\lambda_{\rm max}})$ $\forall$
$\bm{k}$. Hence, there is in this case an exact  {\it universal optimizing
measurement,} determined by the largest semi-axis of the correlation ellipsoid.

Let us now consider the weakly correlated regime. In the two-qubit case,
Eqs.\ (\ref{Sfg}) and (\ref{Hdef}) lead to
\begin{equation}
S_f(A|B_{\bm{k}})\approx h_f(|\bm{r}_A|)-
\frac{\bm{k}^T C^T \Lambda_f(\bm{r}_A)C\bm{k}}{\bm{k}^T N_B\bm{k}}\,,
\label{Sfp}
\end{equation}
where the Hessian matrix  (\ref{Hfg}) becomes now a $3\times 3$ matrix that
depends just on $\bm{r}_A$ and can be expressed as
\begin{eqnarray}
\Lambda_f(\bm{r}_A)
    &=&-\frac{h'_f(r_A)}{2r_A}
\left[I+[\eta_f(r_A)-1]\frac{\bm{r}_A\bm{r}_A^T}{r_A^2}\right]\label{H/h}\,,\label{tql}\\
\eta_f(r)&=&rh_f''(r)/h_f'(r)\,,\label{eta}
\end{eqnarray}
where $r_A=|\bm{r}_A|$. It is then verified that for $f$ concave,
$\Lambda_f(\bm{r}_A)$ is a positive definite $3\times 3$ matrix, since
$\eta_f(r)>0$.

For  the quadratic entropy, $h_f(r)=\frac{1-r^2}{2}$ and $\eta_f(r)=1$,
implying  $\Lambda_f(\bm{r}_A)=I$. It is also verified that for
$\bm{r}_A\rightarrow \bm{0}$ and arbitrary $S_f$, $h_f'(r_A)\rightarrow 0$,
with $h_f'(r_A)/r_A\rightarrow h_f''(0)$ and $\eta_f(r_A)\rightarrow 1$,
implying  $\Lambda_f(\bm{0})=\frac{1}{2}|h''_f(0)|I$, in agreement with
(\ref{Hra0}).  In this limit $\Delta S_f(A|B_{\bm{k}})\approx
\frac{1}{2}|h_f''(0)|\Delta S_2(A|B_{\bm{k}})$ $\forall$ $S_f$ in the
approximation (\ref{Sfp}).

\begin{figure}[htp]
\centering
\includegraphics[scale=0.7]{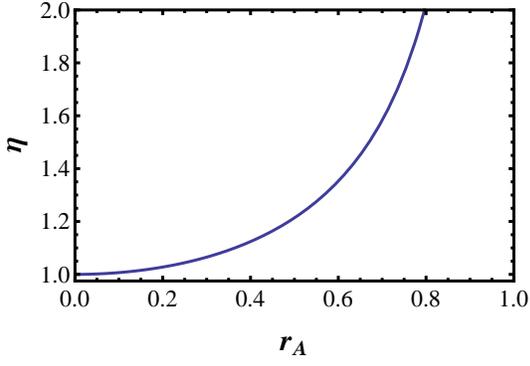}
\caption{(Color online) Plot of the factor $\eta(r_A)=\frac{r_A
h''(r_A)}{h'(r_A)}$ in Eq.\ (\ref{tql}) for the von Neumann entropy. Since it
is an increasing function, differences with the quadratic entropy results (for
which $\eta(r_A)=1$ $\forall$ $r_A$) will increase as
$r_A=|\langle\bm{\sigma}_A\rangle|$ increases. Quantities plotted are
dimensionless.} \label{f3}
\end{figure}

However, for a general $\bm{r}_A$, $\Lambda_f(\bm{r}_A)$  will introduce an
anisotropy in the direction of $\bm{r}_A$ whenever $\eta_f(r_A)\neq 1$. This
factor is  a local measure of the concavity of $h_f$ in the direction of
$\bm{r}_A$, taking as reference the quadratic entropy, and  will favor the
$\bm{r}_A$ direction if $\eta_f(r_A)>1$. This occurs in the Von Neumann case
(Fig.\ \ref{f3}),
where $h_f(r)=h(r)=-\sum_{\nu=\pm 1}\frac{1+\nu r}{2}\log_2\frac{1+\nu r}{2}$
and $\eta_f(r)=\eta(r)$, with
\begin{equation}
\eta(r)=\frac{2r}{(1-r^2)\ln\frac{1+r}{1-r}}>1\,,
\label{etavN}
\end{equation}
for $r>0$ ($\eta(r)\approx 1+2r^2/3$ for $r\rightarrow 0$). However,
$\eta_f(r)<1$ is also possible for a general concave $f$. For instance, for the
Tsallis entropies \cite{TS.09} $S_q(\rho)=(1-{\rm Tr}\,\rho^q)/c_q$, obtained
for  $f(p)=(p-p^q)/c_q$, with $c_q=1-2^{1-q}$ and $q>0$,
\[\eta_q(r)=\frac{(q-1)r}{1+r}\frac{1+\gamma^{q-2}}{1-\gamma^{q-1}}\,,\]
where $\gamma=\frac{1-r}{1+r}$. This leads to $\eta_q(r)>1$ for $q\in(0,2)$ or
$q>3$ but  $\eta_q(r)<1$ for $q\in(2,3)$, with $\eta_q(r)=1$ for $q=2$ or $q=3$
($\eta_q(r)\approx 1+\frac{(q-2)(q-3)}{3}r^2$ for $r\rightarrow 0$). Note that
$S_q(\rho)$ becomes the von Neumann entropy for $q\rightarrow 1$ and  the
quadratic entropy (\ref{S2}) for $q=2$, coinciding again with $S_2(\rho)$ for
$q=3$ in the single qubit case \cite{RCC.10}.

We can now easily understand the main features of the projective measurement
minimizing $S_f(A|B_{\bm{k}})$ for a general $S_f$. For maximally mixed
marginal states, correlation access depends solely on the correlation tensor,
and the maximum correlation direction, i.e., the major axis of the correlation
ellipsoid, is preferred $\forall$ $S_f$. This preference is affected by a non
zero value of $\bm r_B$, which introduces an anisotropic normalization on the
measurement vectors and entails the replacement of $C$ by $C_N=CN_B^{-1/2}$,
which will favor the direction of $\bm{r}_B$.  Finally, for $\bm{r}_A\neq
\bm{0}$ the local concavity induces an additional $f$-dependent anisotropy
around the direction of $\bm{r}_{A}$, which in the weakly correlated regime
amounts to replace $C_N$ by $\sqrt{\Lambda_f(\bm{r}_A)}C_N$. For
$r_B\rightarrow 1$ or in the pure state limit, the approximation (\ref{Sfp})
will normally  break down, since the correlation ellipsoid will typically
become large.

\begin{figure}[htp]
\centering
\includegraphics[scale=0.6]{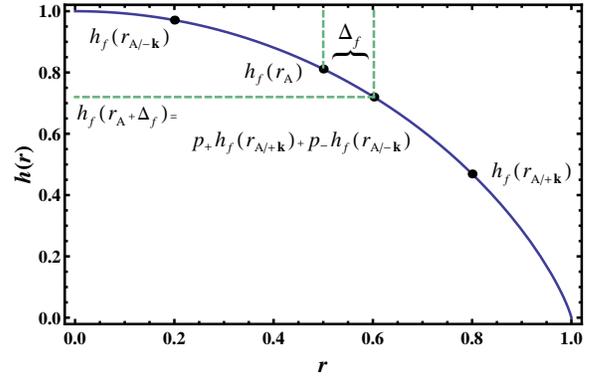}
\caption{(Color online) A measurement is performed in the direction of vector
$\bm{k}$ on qubit $B$ and the entropy of the post-measurement state of $A$ as
measured by $S_f$ is $h_f(|\bm{r}_{A/\pm\bm{k}}|)$. The \emph{measurement
equivalent} $\Delta_f$ is defined as the increase in the norm of vector
$\bm{r}_A$ that satisfies
$h_f(\bm{r}_A+\Delta_f)=p_{\bm{k}}h_f(|\bm{r}_{A/\bm{k}}|)+
p_{-\bm{k}}h_f(|\bm{r}_{A/-\bm{k}}|)$ ($h$ and $r$ dimensionless).} \label{f4}
\end{figure}

{\it Measurement equivalent.} In a two-qubit system, the conditional entropy
decrease at $A$ due to a measurement on $B$ can be characterized by an
effective Bloch vector length increase $\Delta_f$ at $A$, which we will denote
as \emph{measurement equivalent}.  For a projective measurement, it is defined
by (Fig.\ \ref{f4})
\begin{equation}
h_f(|\bm{r}_A|+\Delta_f)=S_f(A|B_{\bm{k}})\,.
\label{Deltaf_def}
\end{equation}
Since $S_f(A|B_{\bm{k}})\leq S_f(A)$, $\Delta_f\geq 0$ for $f$ concave,
increasing as  $\bm{k}$ approaches the optimal direction.

In the weakly correlated regime, $\Delta_f$ will be small. If
$\bm{r}_A\neq\bm{0}$, we then have $h_f(r_A+\Delta_f)\approx
h_f(r_A)+h'_f(r_A)\Delta_f$, and Eq.\ (\ref{Sfp}) leads to $\Delta_f$ of order
$||C_N||^2$:
\begin{eqnarray}
\Delta_f&\approx& \frac{1}{|h'_f(r_A)|}
\frac{\bm{k}^TC^T \Lambda_f(\bm{r}_A)C\bm{k}}{\bm{k}^T N_B\bm{k}}
\label{deltaf1}\,\;\;(r_A>0).
\end{eqnarray}
On the other hand, if $\bm{r}_A\rightarrow \bm{0}$, $h'_f(0)=0$ and  we have
instead $h_f(\Delta_f)\approx h_f(0)+\frac{1}{2}h''_f(0)\Delta^2_f$. Since
$\Lambda_f(\bm{0})=\frac{1}{2}|h''_f(0)| I$, Eq.\ (\ref{Sfp}) leads in this
case  to
\begin{eqnarray}
\Delta_f&\approx& \sqrt{\frac{\bm{k}^TC^T C\bm{k}}{\bm{k}^T
N_B\bm{k}}}=\frac{|C\bm{k}|}{\sqrt{1-(\bm{r}_B\cdot\bm{k})^2}}\,\;\;(\bm{r}_A=0).
 \label{deltaf2}\end{eqnarray}
Thus, for $\bm{r}_A\rightarrow \bm{0}$, $\Delta_f$ becomes {\it independent} of
$f$ ({\it universal limit})  and  of order $||C_N||$.

\section{Application}

\subsection{Quantum discord estimation \label{D}}

Given a bipartite quantum state $\rho_{AB}$ with marginal states $\rho_{A(B)}$,
the quantum discord for a local measurement on $B$ can be written as
\cite{OZ.01}
\begin{eqnarray}
D(A|B)&=&\mathop{\rm Min}_{\{\Pi_j\}}D(A|B_{\{\Pi_j\}}),
\label{discord}\\
D(A|B_{\{\Pi_j\}})&=&S(A|B_{\{\Pi_j\}})-[S(\rho_{AB})-S(\rho_{B})]
 \label{discord0}\end{eqnarray}
where $S(A|B_{\{\Pi_j\}})$ is the conditional entropy (\ref{EC}) in the von
Neumann case, with the minimum in (\ref{discord}) taken over all possible
measurements on $B$, while the bracket in (\ref{discord0}) is the measurement
independent quantum conditional entropy. We may also rewrite (\ref{discord0}) as
\begin{equation}
D(A|B_{\{\Pi_j\}})=I(A,B)-\Delta S(A|B_{\{\Pi_j\}})\,,
\label{discord2}
\end{equation}
where $\Delta S(A|B_{\{\Pi_j\}})=S(A)-S(A|B_{\{\Pi_j\}})$ and
\begin{equation} I(A,B)=S(\rho_A)+S(\rho_{B})-S(\rho_{AB})\,,\label{IAB}
\end{equation}
is the quantum mutual information.

For qudit-qubit systems, the results of previous section can be applied to
estimate Eqs.\ (\ref{discord})-(\ref{discord2}) in the weakly correlated
regime. For a projective measurement along direction $\bm{k}$ at $B$,  Eq.\
(\ref{deltaf}) leads to
\begin{equation}
D(A|B_{\bm k})\approx I(A,B)-\frac{2}{d_A}\frac{\bm{k}^TC^T
\Lambda(\rho_A)C\bm{k}}{\bm{k}^TN_B\bm{k}},\label{disc2}
\end{equation}
where $\Lambda(\rho_A)$ is the Hessian matrix (\ref{Hfg}) in the von Neumann
case. The minimization in (\ref{discord}) leads  then to the eigenvalue problem
(\ref{eigg}), and the minimum reads
\begin{equation}
D(A|B)\approx I(A,B)-\frac{ 2}{d_A}\lambda_{\rm max}, \label{Dap}\end{equation}
with $\lambda_{\rm max}$ the largest root of ${\rm Det}[C^T\Lambda(\rho_A)
C-\lambda N_B]=0$. While $I(A,B)$ is a measure of the total correlation between
$A$ and $B$, the second term in (\ref{Dap}) represents the maximum
classical-like mutual information obtained after a local measurement on $B$, in
the present regime.

In this regime we may also apply a quadratic approximation to (\ref{IAB}) using
the representation (\ref{rhoab}) of $\rho_{AB}$. An expansion of $S(\rho_{AB})$
up to second order in the correlation tensor $C$, extending Eqs.\
(\ref{exp1})--(\ref{exp2}) to this case ($|i\rangle\rightarrow |i_Aj_B\rangle$,
$p_i^A\rightarrow p_{i}^A p_j^B$, $\delta
\rho_A\rightarrow\delta\rho_{AB}=\rho_{AB}-\rho_A\otimes\rho_B$,
 with
$\rho_{A(B)}|i_{A(B)}\rangle=p_i^{A(B)}|i_{A(B)}\rangle$), leads to
\begin{eqnarray}
I(A,B)&\approx&\frac{1}{2}\bm{C}^T\Lambda(\rho_A,\rho_B)\bm{C}\,,
\label{CCH}\end{eqnarray}
where $\bm{C}$ denotes a vector of elements $C_{\mu\nu}$ and $\Lambda$ is here
the $3D_A\times 3D_A$ matrix
\begin{eqnarray}
\Lambda_{\mu\nu}^{\mu'\nu'}(\rho_A,\rho_B)&=&{\textstyle\frac{1}{4d_A^2}\sum
\limits_{i,j,k,l}R_{ik}^{jl}
\langle i_A|\sigma_{A\mu'}|j_A\rangle\langle j_A|\sigma_{A\mu}|i_A\rangle}\nonumber\\
&&{\times \langle k_B|\sigma_{B\nu'}|l_B\rangle\langle
l_B|\sigma_{B\nu}|k_B\rangle}\,, \label{HH}\end{eqnarray}
\begin{eqnarray}R_{ik}^{jl}&=&{\textstyle\frac{1}{\ln 2}[(1-\delta_i^j\delta_k^l)
\frac{\ln[p_i^Ap_k^B/(p_{j}^A p_{l}^B)]}{p_i^Ap_k^B-p_j^Ap_l^B}+\delta_i^j\delta_k^l
\frac{1}{p_i^A p_k^B}}]\,.\nonumber
\end{eqnarray}
The terms linear in $C$ vanish for the von Neumann entropy. Eq.\ (\ref{disc2})
becomes then a quadratic form in the elements of the correlation tensor, which
is positive semidefinite since $D(A|B)\geq 0$ and the quadratic approximation
becomes  exact for sufficiently small $C$.

The decomposition (\ref{CD}) allows to reduce Eq.\ (\ref{CCH}) to a quadratic
form in the three singular values $C_\mu$. For instance, for maximally mixed
marginals, $R^{jl}_{ik}=2d_A/\ln 2$ $\forall$ $i,j,k,l$, implying
$\Lambda^{\mu\nu}_{\mu'\nu'}(\frac{I_A}{d_A},\frac{I_B}{2})=\frac{1}{\ln
2}\delta^\mu_{\mu'}\delta^{\nu}_{\nu'}$. Eq.\ (\ref{CCH}) then reduces to
$I(A,B)\approx \frac{1}{2\ln 2} {\rm Tr}\,C^TC=\frac{1}{2\ln 2}\sum_{\mu=1}^3
C_\mu^2$.

\subsection{Two-qubit states with  $\bm{r}_A$ and $\bm{r}_B$
parallel to a principal axis of $C$}

In the special two-qubit case where  $\bm{r}_A$ and $\bm{r}_B$ are parallel to
one of the principal directions selected by $C$ in the diagonal representation
(\ref{CD}) (implying that they should be eigenvectors of $CC^T$ and $C^TC$
respectively), tensors $\Lambda_f(\bm{r}_A)$, $C$ and
$N_B=I-\bm{r}_B\bm{r}_B^T$ can be made {\it simultaneously diagonal}: We may
choose the local orthogonal $x,y,z$ axes at $A$ and $B$ such that for
$\mu,\nu=x,y,z$,
\begin{equation}
C_{\mu\nu}=\delta_{\mu\nu} C_\mu\,,\label{Cd}
\end{equation}
with $C_\mu$ the singular values of $C$ and $\bm{r}_A$ and $\bm{r}_B$ parallel
to one of these axes. Eqs.\ (\ref{eig2}) and (\ref{eigg}) then imply that the
optimal  measurement minimizing the conditional entropy in the
weakly correlated regime (and in all cases for the quadratic entropy)  is to
be found {\it among these principal axes}.

If $\bm{r}_A$ and $\bm{r}_B$ are both directed along $\bm{z}$
(i.e., $\bm{r}_A\propto C\bm{r}_B$),
Eqs.\  (\ref{deltaf}), (\ref{tql}) and (\ref{Cd}) lead to
\begin{equation}
\Delta S_f(A|B_{\bm{k}})\approx
{\textstyle\frac{|h'_f(r_A)|}{2r_A}\,\frac{C_{x}^2 {k}_{x}^2+C_y^2 {k}_{y}^2+
\eta_f(r_A)C_z^2 k_{z}^2}{1-r_B^2k_z^2}}\,,\label{Szz}
    \end{equation}
with its maximum then given by
\begin{equation}
\mathop{\rm Max}_{\bm{k}}\Delta S_f(A|B_{\bm{k}})\approx
{\textstyle\frac{|h'_f(r_A)|}{2r_A}{\rm Max}[C_x^2,C_y^2,
\frac{\eta_f(r_A)}{1-r_B^2}C_z^2]}\,.\label{Szzm}
\end{equation}

On the other hand, if $\bm{r}_A$ and $\bm{r}_B$ are along orthogonal principal
axes ($\bm{r}_A\perp C\bm{r}_B$), for instance $\bm{r}_B$ along $\bm{z}$ and
$\bm{r}_A$ along $\bm{x}$,  we obtain instead
\begin{equation}
\Delta S_f(A|B_{\bm{k}})\approx {\textstyle\frac{|h'_f(r_A)|}{2r_A}
\,\frac{\eta_f(r_A)C_x^2 {k}_{x}^2+C_y^2 {k}_{y}^2+C_z^2k_z^2}
{1-r_B^2 k_z^2}}\,,\label{Sxz}
\end{equation}
with its maximum given by
\begin{equation}
\mathop{\rm Max}_{\bm{k}}\Delta S_f(A|B_{\bm{k}})\approx
{\textstyle\frac{|h'_f(r_A)|}{2r_A}
{\rm Max}[\eta_f(r_A)C_x^2,C_y^2,\frac{C_z^2}{1-r_B^2}]}\,.\label{Sxzm}
\end{equation}
For use in the next subsection, we quote here the explicit expressions for the
case of the von Neumann entropy when $\bm{r}_A$ and $\bm{r}_B$ are both
parallel to $\bm{z}$. We obtain
\begin{equation}
\Delta S(A|B_{\bm{k}})\approx {\textstyle\frac{\frac{1}{2r_A}
\ln\frac{1+r_A}{1-r_A} (C_x^2 {k}_x^2+C_y^2 {k}_y^2)+
\frac{1}{1-r_A^2}C_z^2k_z^2}{2\ln 2 \,(1-r_B^2k_z^2)}}
\,,\label{Szza}
    \end{equation}
whereas the quadratic approximation (\ref{CCH}) becomes
\begin{eqnarray}
I(A,B)&\approx&{\textstyle\frac{1}{2\ln 2}}[\!{\textstyle
\sum\limits_{\nu=\pm 1}\!\frac{(C_x-\nu C_y)^2
\ln(\frac{1+r_A}{1-r_A}\,\frac{1+\nu r_B}{1-\nu r_B})}{4(r_A+\nu r_B)}}
+{\textstyle\frac{C_z^2}{(1-r_A)^2(1-r_B)^2}]}\,.\nonumber\\&&\label{Ia}
\end{eqnarray}
It is verified that within the approximations (\ref{Szza})--(\ref{Ia}),
$D(A|B_{\bm{k}})=I(A,B)-\Delta S(A|B_{\bm{k}})$ becomes a non-negative quadratic
form in the $C_\mu$'s.

\subsection{Optimum measurement for $X$ states \label{sX}}
We now apply previous approximations to the set of two-qubit $X$ states,
which arise naturally in many physical situations \cite{BL.08,RD.08,CRC.10}.
Through the singular value decomposition of the tensor $J_{\mu\nu}=\langle
\sigma_\mu\otimes\sigma_\nu\rangle$, and by suitably choosing the local bases,
these states can be written as
\begin{eqnarray}
\rho_{AB}&=&\frac{1}{4}(I\otimes I+r_A\sigma_z\otimes I
+r_B I\otimes \sigma_z+\sum_{\mu}
 J_\mu \sigma_\mu\otimes\sigma_\mu) \label{X1}
\\
&=&\left(\begin{array}{cccc}p_+&0&0& \alpha_-\\0&q_+& \alpha_+&0\\
0&\alpha_+ & q_-& 0\\\alpha_-&0&0&p_-\end{array}\right);
\begin{array}{c}
p_{\pm}=\frac{1\pm(r_A+r_B)+J_z}{4}\\
q_{\pm}=\frac{1\pm(r_A-r_B)-J_z}{4}\\
\alpha_{\pm}=\frac{J_x\pm J_y}{4}\end{array}\label{X}
\end{eqnarray}
with (\ref{X}) the state representation in the standard basis. The parameters
should fulfill the positivity conditions $p_{\pm}\geq 0$, $q_{\pm}\geq 0$,
$|\alpha_-|\leq \sqrt{p_+p_-}$, $|\alpha_+|\leq \sqrt{q_+q_-}$, with
 $p_++p_-+q_++q_-=1$. Since the correlation tensor
$C=J-\bm{r}_A\bm{r}_B^T$ will satisfy Eq.\ (\ref{Cd}), with
\[C_{x}=J_{x}\,,\;\;\;C_y=J_y\,,\;\;\;C_z=J_z-r_A r_B\,,\]
it is clear that in these states the marginal Bloch vectors $\bm{r}_A$ and
$\bm{r}_B$ lie on the same principal axis ($z$) of $C$,  implying that
these states correspond to the case of Eq.\ (\ref{Szz}). In the weakly
correlated limit $S_f(A|B_{\bm{k}})$ will then reach its minimum  for a
measurement along the direction of one of these principal axes (Eq.\
(\ref{Szzm})).

In this regime the minimizing measurement depends not only on  $C$ and $r_B$,
but also on the local concavity of the function $h_f(r)$ at $r=|\bm{r}_A|$.
This implies, in general, that different entropies may reach their minimum
value for measurements on different axes. We will now compare the minimizing
measurements of the von Neumann and quadratic conditional entropies for states
with $J_x=J_y$, for  which the minimizing measurement is either along the $z$
axis or along any vector in the $x,y$ plane, which we will take as $x$. A
\emph{transition zone} between these two directions arises, that will depend on
the concavity of the entropy. From Eq.\ (\ref{Szzm}) it follows that  the
transition zone is
\begin{equation}
C_x^2=\eta_f(r_A)C_z^2/(1-r_B^2),\label{transz}
\end{equation}
with $\eta_f(r_A)=1$ for the quadratic entropy and $\eta_f(r)=\eta(r)$, Eq.\
(\ref{etavN}), for the von Neumann entropy. Since $\eta(r_A)>1$ for $r_A\neq
0$, it is seen that in the von Neumann case, the transition zone is shifted
from that of the quadratic entropy whenever $r_A\neq 0$, and this discrepancy
will increase as $r_A$ increases, favoring the $z$ direction.

\begin{figure}[htp]
\centering
\includegraphics[scale=0.6]{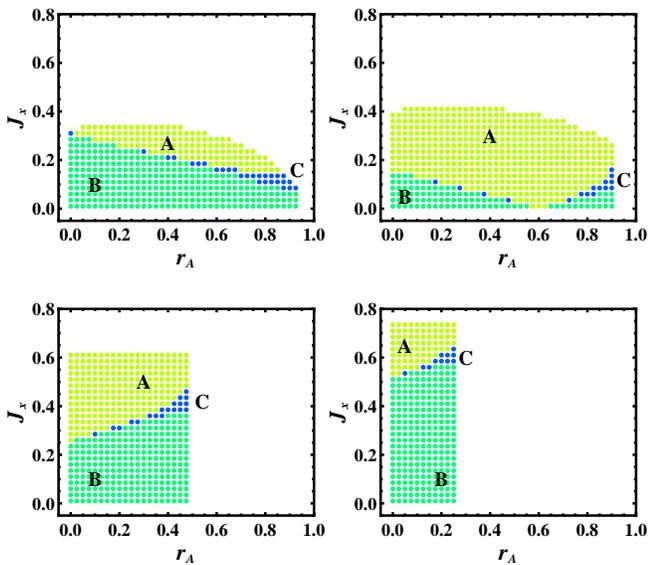}
\caption{(Color online) Comparison between the projective minimizing
measurements for the von Neumann and quadratic conditional entropies, for $X$
states (Eq.\ (\ref{X1})) with $r_B=0.25$ and $J_z=0.3$ (top left), 0.15 (top
right), -0.25 (bottom left), -0.5 (bottom right). Yellow (sector A) and green
(sector B) disks show the set of states where the minimizing measurement is the
same for both entropies (along $\bm{x}$ in A and along $\bm{z}$ in B), while
blue disks (C) show those states where the minimizing measurements differ
($J_x$ and $r_A$ dimensionless).} \label{f5}
\end{figure}

Typical results for the projective minimizing measurement for these entropies
are shown in Fig.\ \ref{f5} as a function of $r_A$ and $J_x=C_x$, for fixed
$r_B$ and different values of $J_z$. It is seen that they are coincident for
most states, differing only in the transition region C (blue disks), where
the measurement minimizing the quadratic entropy has already changed from $\bm
z$  to the $\bm x$ direction, but the von Neumann entropy still reaches its
minimum value for a measurement along $\bm z$. As expected, the region of
discrepancy becomes greater as $r_A$ increases. We should mention that while
the $\bm{z}\rightarrow \bm{x}$ transition as $J_x$ increases is always sharp
for the quadratic conditional entropy, as follows from Eq.\ (\ref{S2x}), in the
von Neumann case  it may be softened through intermediate measurement
directions in a tiny interval of $J_x$ values, an effect not seen in the
approximation (\ref{Szz})--(\ref{Szza}).  Actually, in these tiny crossover
intervals  non-projective measurements can be preferred \cite{CZ.11,KS.13} (if
a projective measurement optimizes the von Neumann conditional entropy for an
$X$ state, it should be along a principal axes of $C$ \cite{KS.13}), although
differences with the projective minimum are small.

\begin{figure}[htp]
\centering
\includegraphics[scale=.7]{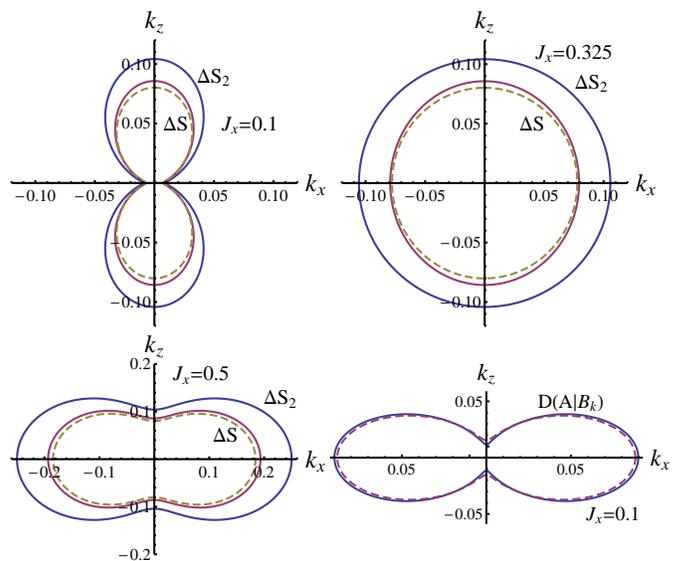}
\caption{(Color online) The entropy decrease $\Delta S_f=S(A)-S(A|B_{\bm{k}})$
after a local measurement on $B$ along direction $\bm{k}=(k_x,0,k_z)$ for the
quadratic ($\Delta S_2$) and von Neumann ($\Delta S$)  entropies, together with
the quadratic approximation (\ref{Szz})--(\ref{Szza}) for the latter (dashed
lines). We have considered an $X$ state with $r_A=r_B=0.25$, $J_z=-0.25$ and
$J_x=0.1, 0.325, 0.5$, corresponding to states in sectors B, C, and A,
respectively, of the lower left panel of Fig.\ \ref{f5}. The bottom right panel
depicts the quantum discord (\ref{discord2}) and
its quadratic approximation (\ref{Szza})--(\ref{Ia}) (dashed line) for
$J_x=0.1$ (quantities plotted dimensionless). } \label{f6}
\end{figure}

Fig.\ \ref{f6} shows  the entropy decrease (``information gain'') $\Delta
S_f=S_f(A)-S_f(A|B_{\bm{k}})$ as a function of the direction
$\bm{k}=(k_x,0,k_z)$ of the measurement on $B$, for $X$ states located below, at
and above the transition zone in the bottom left panel of Fig.\ \ref{f5}.  Both
the quadratic and von Neumann conditional entropies are depicted, which are
seen to exhibit typically the same profile, together with the second order
approximation (\ref{Szza}) to the latter, which is seen to provide a good
estimation. While there is a clear preference for the $\bm{z}$ ($\bm{x}$)
direction for low (high) $J_x$, the anisotropy of $\Delta S_f$ in the
transition region ($J_x=0.325$), where the minimizing measurement
directions of the von Neumann and quadratic entropies differ, is very small,
entailing that this difference is not too relevant.  We also depict
illustrative results for the discord (\ref{discord2}) and its quadratic
estimation obtained with Eqs.\ (\ref{Szza})--(\ref{Ia}), which is quite accurate
in the case considered.

\section{Conclusions}

We have shown that the problem of conditional entropy optimization in a
qudit-qubit system, for a general entropic form and a measurement on the qubit,
can be solved analytically in the limit of weak correlations. It just requires
the solution of a $3\times 3$  eigenvalue problem determined by the correlation
tensor of the system, the Bloch vector of the qubit and a local concavity term
depending on the choice of entropy (Eqs.\ (\ref{Sfg}), (\ref{eigg})).  In the
case of the quadratic entropy, which is directly related to the purity (and is
hence experimentally accessible without requiring a full state tomography
\cite{RF.02}), the concavity term reduces to an identity matrix and the
approach is exact in all regimes. The optimization problem admits in this case
a direct geometrical interpretation in terms of the correlation ellipsoid
representing the set of post-measurement states of the qudit, with the
minimizing measurement direction determined by its largest principal axis,
i.e., by the direction which optimizes correlation access.

For a general entropic form, the corrections for a sufficiently small
correlation ellipsoid lead to the effective correlation tensor (\ref{rep}),
which includes the effects of the local ``concavity excess''  through a Hessian
matrix. This allows first to identify some universal features of the problem,
such as the common (valid for all entropies) profile and minimizing measurement
in this regime when the marginal state of the qudit is maximally mixed. When
applied to the von Neumann entropy, the present scheme also leads to a simple
direct estimation of the quantum discord, including a fully quadratic (in the
correlation tensor) approximation  after a concomitant expansion of the mutual
information. Illustrative results for two qubit $X$ states indicate a good
agreement of the present approximations with the exact values beyond the very
weak correlation limit, with similar profiles for the quadratic and von Neumann
entropy in typical situations. Application of the present approach to more
complex many body systems and measures are presently being considered.

The authors acknowledge support of CIC (RR) and CONICET (NG) of Argentina.

\end{document}